# Transportation Emergency Planning Considering Uncertainty in Event Duration and Drivers' Behavior

Fardad Haghpanah[*]

*School of Civil, Environmental and Land Management Engineering, Politecnico di Milano, Milan, Italy*

**Abstract**

Traffic Emergency Management deals with directing the vehicular and pedestrian traffic around traffic disruptions due to emergencies, such as accidents or flooded roadways, aiming to ensure the safety of drivers, pedestrians, and emergency responders. In this study, a scenario involving the local flooding of the A1 motorway, one of Italy's main highways connecting north to the south, is studied. The effect of event duration and drivers' response rate are investigated on the alternative route activation strategies. The macro and micro itineraries are established, and for different event durations and response rates, the timelines for effective route activation are evaluated. According to the results, for events shorter than 1.5 hours, there is no need for the activation of alternative routes, and the longer the event, the more alternative routes are needed to minimize the total travel time on the flooded route. In addition, increase in the response rate of drivers to use the alternative routes leads to the need to activate the micro itinerary after the activation of the macro itinerary. Furthermore, the evacuation of an urban region due to the flood scenario is studied considering different evacuation strategies and residents response time. The results indicate the importance of optimal exit point allocation and residents' preparedness to reduce the total evacuation time.

## 1. Introduction

Traffic management mainly focuses on ensuring the users to receive a desired quality of service. This could be challenging task during periods of heavy traffic loads or in case of road network failures such as flooded roadways. As a result, congestion control is the most essential aspect of traffic management [1]. Traffic Emergency Management involves directing the vehicular and pedestrian traffic around traffic disruptions such as accidents or flooded roadways, thus ensuring the safety of drivers, pedestrians, and emergency responders. In addition, it includes the use traffic monitoring systems (e.g. CCTV) by the local roadway authorities to manage traffic flows and provide advice regarding traffic congestions. Accordingly, Traffic Emergency Management Plan (TEMP) is a document established to manage traffic disruptions due to emergencies that need coordination from several agencies involved in road and traffic management [2]. It describes the necessary actions to control the traffic in case of an emergency, including alternative routes, criteria for activation and deactivation of the emergency routes, and location of barricades, warning lights, and signs.

The main body of research on traffic emergency management has been devoted to the planning aspects [3], including traffic emergency management policy [4,5], origin-destination trip estimation [6,7], evacuation modeling [8,9], and behavior analysis [10,11]. Furthermore, since each event, i.e. natural disasters or road accidents, has specific characteristics and corresponding preparatory and responsive actions, specific planning models have been developed for different emergency events, such as hurricane, flooding, wildfire, and earthquake. These models are helpful tools for emergency management professionals and policy makers to evaluate different strategies during emergencies, and consequently for societies to move toward building a more sustainable and resilient infrastructure [12,13].

In this study, a scenario involving the local flooding of the A1 motorway, one of Italy's main highways connecting north to the south, is studied. Multiple transportation emergency strategies are developed considering different event durations and drivers' response rates. The macro-itineraries (including alternative highways) and alternative micro itinerary (including local road network only) are obtained using ArcMap Network Analyst package and a transportation and land-use modeling platform, the Citilabs Cube. Furthermore, the evacuation of an urban region due to the flood scenario is studied considering different evacuation strategies and residents response time.

*\* Current address: Department of Civil Engineering, Johns Hopkins University, Baltimore, MD 21218, USA*
*Email: haghpanah@jhu.edu*

## 2. Study Area

Po River passes through the boundary of Lombardy and Emilia-Romagna regions in Italy and is subjected to heavy flooding, specifically the southern part of the Lodi area adjacent to the left bank of the river which is located below the terrace morphology.

This is a highly populated area that in case of flooding, would suffer substantial damage. The area is crossed by the most important arterials (roads and railways) linking the north and south of Italy. These arterials will be dropped in case of an extreme emergency event. The major communication routes that cross Lodi, the flood-prone area, are the A1 motorway, the High Speed Rail, the Statal Road SS9 (Via Emilia), and the Milan-Bologna railway. The Autostrada A1 (known also as Autostrada del Sole, literally "Motorway of the Sun") is an Italian motorway that connects Milan with Naples via Bologna, Florence, and Rome. At 754 km, it is the longest Italian autostrada and is considered the spinal cord of the country's road network. It is also a part of the E35 and E45 European roads.

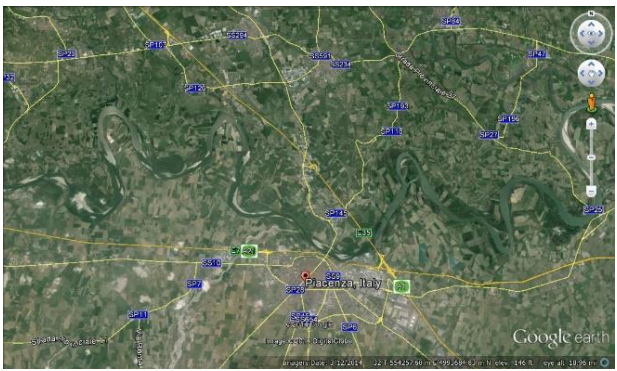

Figure 1: Po River and the study area (source: Google Map)

### 2.1. Transportation Network and Traffic Data

A set of GIS shapefiles is provided including the administrative boundaries, municipalities and provinces, centroids, connectors, highway road network, local road network, flood risk regions, and location of variable message signs on highways. In addition, the national demand matrix including 100 nodes (90 provinces and 10 cross-border centroids) is provided.

## 3. Modeling and Analysis

The aim of this study is to manage the traffic on the A1 motorway during a flood scenario. To manage the disrupted traffic, three types of alternative paths have been considered: (1) the macro itinerary which includes solely the highways and major roads; (2) the micro itinerary which includes only the local road network; and (3) the hybrid itinerary which includes both the major and the local roads. Different strategies to activate the alternative routes are developed based on the duration of the flooding event and the response rate of drivers to the instructions. Furthermore, the evacuation procedure of the township of San Rocco al Porto, located above the Po River, due to the flood scenario is evaluated. The total evacuation time is estimated for three different shelter allocation and for two residents' departure time settings, resulting in six evacuation scenarios.

In order to model the road network and analyze the traffic flow through the network and alternative paths considering the flood scenario, Citilabs Cube is used. Cube is a transportation and land-use modeling platform with many applications in transportation management and includes the most complete suite of software products consisting of eight modules to form a complete travel forecasting and transportation GIS system. The macro itineraries and the evacuation time are obtained using Cube, and the ArcMap Network Analyst package is used to acquire the micro itinerary.

### 3.1. The Scenario

The scenario considered for this study is the flooding of the Po River, located just to the north of Piacenza, which will lead to the closure of the route A1. The flood occurs on a weekday at 12:00 PM, and the north to south traffic flow on the A1 motorway is considered for analyses. The duration of the flood is variable and is 0.5, 1.5, 2, and 2.5 hours.

In order to obtain the hourly traffic for the scenario, the national demand matrix is used. Assuming that the traffic flow on Saturdays is 67.9% of a standard working day, and is 49.4% for Sundays and holidays (based on local traffic records), considering 51 Saturdays, 51 Sundays, 12 holidays and 251 normal days in a year in Italy, the daily flow for a standard working day will be:

$$Annual\ Flow = 251 \times 1 \times F_{Daily} + 51 \times 0.679 \times F_{Daily}$$
$$+ (51+12) \times 0.494 \times F_{Daily} \quad (1)$$
$$F_{Daily} = \frac{Annual\ Flow}{316.751}$$

Therefore, the hourly traffic flow on a working day at 12:00 PM based on hourly distribution of the traffic (from local traffic records) is:

$$F_{Hourly} = 5.34\%\ F_{Daily} = \frac{Annual\ Flow}{316.751} \times 0.0534 \quad (2)$$

## 4. Results

The results of the traffic distribution using all-or-nothing assignments is shown in Figures 2. Figure 2a illustrates the average national hourly traffic as obtained by Eq. 2. Figure 2b shows the average hourly traffic that passes through the flooded section of motorway A1, which is the traffic to be disrupted by the flood event. Figure 2c shows how the disrupted traffic would be redistributed to other highways by excluding the flooded section of motorway A1 from the national road network and reassigning the traffic. This will help us to establish the macro itinerary to redirect the disrupted flow. In Figure 2d, the blue-to-red lines is the disrupted traffic on the flooded section of motorway A1 and is acquired by subtracting the flow due to the closure of the link from the normal flow through the network. The yellow-to-green lines is the diverted flow and is obtained by subtracting the normal flow through the network from the flow due to the closure of the link.

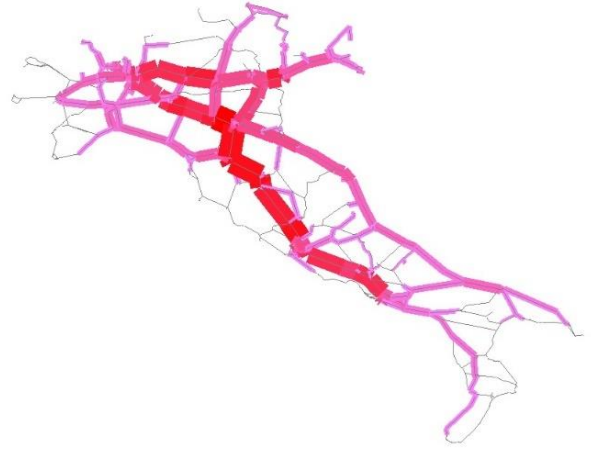

(a)

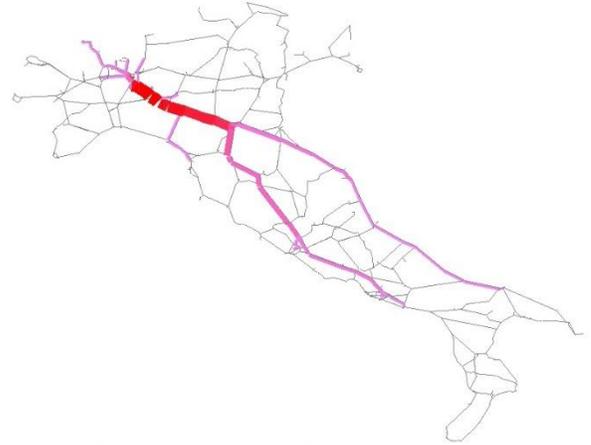

(b)

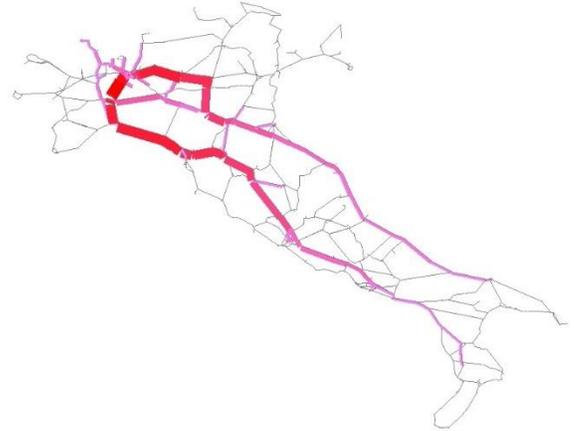

(c)

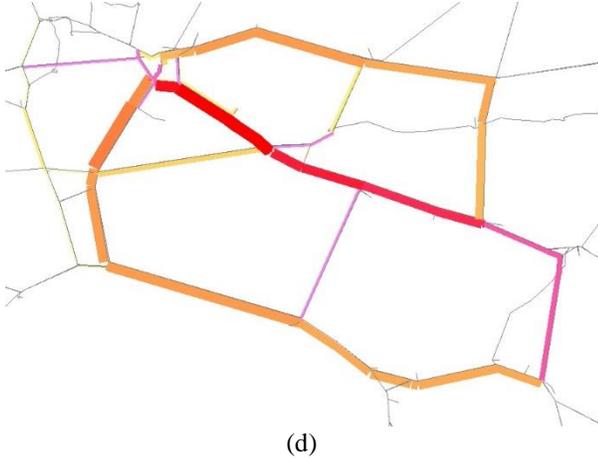

(d)

Figure 2: (a) Average national hourly traffic, (b) Disrupted traffic due to the flooding, (c) Disrupted traffic redistributed to other routes, (d) disrupted and redistributed flows on the flooded section of motorway A1.

### 4.1. The Macro Itinerary

In this section, an alternative route is determined substitute the closed link using only the highway network.

According to the CUBE simulation results, by closing the flooded link, the interrupted flow will take two alternative routes depending on the destinations. It can be interpreted that those whose destinations are west of Italy will take the left route, and those whose destinations are east will take the right ones. Thus, as shown in Figure 3, the two macro alternative routes will have quite the same starting point but their end points are different.

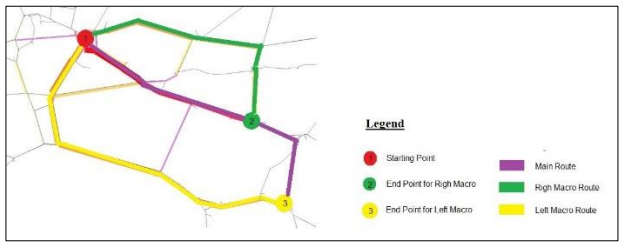

Figure 3: The macro itinerary

After establishing the Macro itineraries, the travel time for each route has to be calculated and compared with the travel time on the main route to evaluate the efficiency of the itineraries. Based on the results, from the total interrupted flow of 3750 vehicle per hour, 1650 vehicle per hour will take the right route, and 2100 vehicle per hour will take the left one. Furthermore, initially it is assumed that 60% of drivers would response to the instructions, and 40% will decide to stay on the main route. To calculate the actual travel time of the macro itineraries, the routes are divided into segments and the travel time for each segment is calculated as follows [14]:

$$t = \frac{L}{v}, \quad v = v_f - \alpha \left(\frac{q}{L_u n}\right)^2 \geq 5 \ km/h \quad (3)$$

in which, $t$ is travel time, $L$ is the length of the segment, $v$ is the flow speed, $v_f$ is the free-flow speed, $\alpha$ is the model parameter (0.0001), $q$ is the traffic flow (daily flow plus the diverted flow), $L_u$ is the lanes' working width (3.5 m), and $n$ is the number of lanes. The daily (i.e. hourly) flow is obtained from the All or Nothing assignment results from the Cube, and the diverted flow is 60% of the disrupted flow as explained above. The results are listed in Table 1. The delay times due to the road block on the main route will be added to the travel time of the main route for route activation decision.

Table 1: Travel time for the macro itineraries and the main route without delay times

| Travel time (min) | Travel time (min) | Length (km) |
|---|---|---|
| Left macro itinerary | 191.7 | 345.1 |
| Main route wrt left macro flow | 157.3 | 288.3 |
| Right macro itinerary | 138.1 | 231.5 |
| Main route wrt right macro flow | 89.8 | 162.9 |

### 4.2. The Micro Itinerary

In this section, another alternative route is determined to substitute the closed link using only the local road network. To find the micro route, the Network Analyst Tool in ArcMap is used. The start and end points are placed on the motorway A1 just before and after the flooded section and close to the intersection of the local network with the highway network. In addition, point barriers are located on the highway before the flooded area to close the highway and on the local roads passing through the flooded area to avoid entering traffic into the flooded area. In establishing the micro itinerary, it is considered that all the diverted flow is deviated into the micro alternative route. The results are shown in Figure 4. The travel times are obtained as before and listed in Table 2.

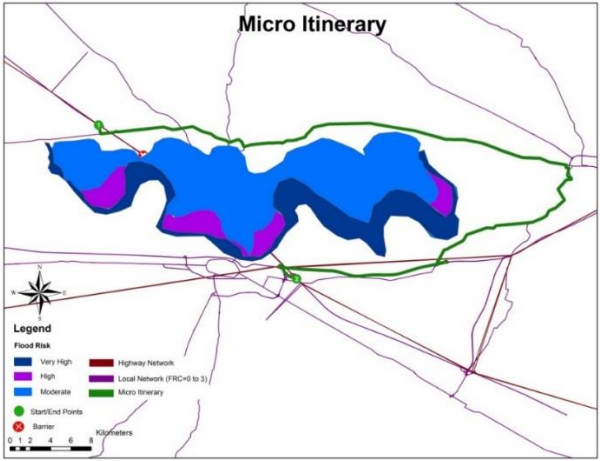

Figure 4: The micro itinerary

Table 2: Travel time for the micro itinerary and the main route without delay times

| Travel time (min) | Travel time (min) | Length (km) |
|---|---|---|
| Micro itinerary | 645.2 | 66.9 |
| Main route wrt micro flow | 11.5 | 20.8 |

### 4.3. Delay time

To estimate the actual travel time on the main route, the delay caused by the event has to be incorporated into the evaluation. To calculate the delay, the input-output diagram method [15] is used. The parameters for the input-output diagram are listed in Table 3.

Table 3: Parameters for the input-output diagram

| Parameters | Values | Units |
|---|---|---|
| Starting time | 12:00 pm | (hr) |
| Flow | See Table 4 | (veh/hr) |
| Capacity ($q_{max}$) | 6000 | (veh/hr) |
| $k_j$ | 600 | (veh/km) |
| $V_f$ | 120 | (km/hr) |
| $k_c$ | 50 | (veh/km) |
| Headway | See Table 4 | (min) |
| $V_{\mu 1}$ | 0.5 | (km/hr) |
| d | 1.0 | (km) |
| $t_f$ | 0.01 | (hr) |
| $\mu_1$ | 1 | (veh/hr) |
| $\mu_2$ | 6000 | (veh/hr) |

Table 4: Flow, headway, and cumulative flow for different time slots during the flooding

| From | To | Flow (veh/hr) | Headway (hr) | Cumulative Flow (veh) |
|---|---|---|---|---|
| 12:00 | 13:00 | 3750.84 | 0.0160 | 3751 |
| 13:00 | 14:00 | 3694.65 | 0.0162 | 7446 |
| 14:00 | 15:00 | 4045.85 | 0.0148 | 11492 |

The resultant delay times are as follows: (1) for event durations up to one hour, the delay is 26.1 minutes; (2) for event durations between one and two hours, the delay is 85.9 minutes; and (3) for event durations between two and three hours, the delay is 145.9 minutes.

### 4.4. Activation of Alternative Routes

For each event duration, the criterion to activate the alternative routes is:

$$T_{alternative\ route} \leq T_{upstream} + T_{downstream} + T_{delay} \quad (4)$$

The activation settings for the micro itinerary, the left macro itinerary, the right macro itinerary, and the micro-macro itinerary are listed in Tables 5, 6, 7, and 8, respectively.

Table 5: Activation of the Micro Itinerary

| Scenario duration (hr) | Main route total travel time (min) | Micro route travel time (min) | Activation of the micro itinerary |
|---|---|---|---|
| 0.5 | 37.6 | 645.2 | No |
| 1 | 37.6 | 645.2 | No |
| 1.5 | 97.4 | 645.2 | No |
| 2 | 97.4 | 645.2 | No |
| 2.5 | 157.4 | 645.2 | No |
| 3 | 157.4 | 645.2 | No |

It is clear that considering only the micro itinerary to be substituted by the main route is not an efficient solution. The reason why the travel time for the micro itinerary is considerably larger than the main route is that all the interrupted flow (3750 veh/hr) is deviated into the local network which mainly constitutes of 1-lane roads with very low section speeds (5 km/hr) due to the congestion.

Table 6: Activation of the left Macro Itinerary

| Scenario duration (hr) | Main route total travel time (min) | Micro route travel time (min) | Activation of the micro itinerary |
|---|---|---|---|
| 0.5 | 183.4 | 191.7 | No |
| 1 | 183.4 | 191.7 | No |
| 1.5 | 243.2 | 191.7 | Yes |
| 2 | 243.2 | 191.7 | Yes |
| 2.5 | 303.2 | 191.7 | Yes |
| 3 | 303.2 | 191.7 | Yes |

Table 7: Activation of the right Macro Itinerary

| Scenario duration (hr) | Main route total travel time (min) | Micro route travel time (min) | Activation of the micro itinerary |
|---|---|---|---|
| 0.5 | 115.9 | 138.1 | No |
| 1 | 115.9 | 138.1 | No |
| 1.5 | 175.7 | 138.1 | Yes |
| 2 | 175.7 | 138.1 | Yes |
| 2.5 | 235.7 | 138.1 | Yes |
| 3 | 235.7 | 138.1 | Yes |

Therefore, for scenario durations longer than 1.5 hours, it is efficient to activate both the left and right macro itinerary.

In case of activating the macro itineraries, according to the hypothesis, 60% of the flow would be diverted from the main route, hence with just 40% of the flow remaining on the main route, the criteria to activate the micro itinerary could change.

Table 8: Activation of the Macro-Micro Itinerary

| Scenario duration (hr) | Main route total travel time (min) | Micro route travel time (min) | Activation of the micro itinerary |
|---|---|---|---|
| 0.5 | 23.6 | 111.2 | No |
| 1 | 23.6 | 111.2 | No |
| 1.5 | 62.8 | 111.2 | No |
| 2 | 62.8 | 111.2 | No |
| 2.5 | 122.0 | 111.2 | Yes |
| 3 | 122.0 | 111.2 | Yes |

Based on the results, activation of the micro itinerary after activating the macro itinerary is efficient only for scenario durations longer than 2.5 hours.

In summary, according to the different alternative routes' activation times, there are 3 scenarios for itineraries to be opened based on the event duration. Figure 5 demonstrates different strategies for the activation of alternative itineraries.

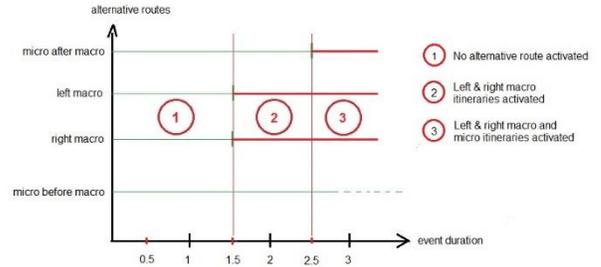

Figure 5: Strategies for the activation of alternative itineraries

### 4.5. Sensitivity Analysis: drivers' response rate

The results of the alternative routes activation are dependent on several factors. An important factor is the percentage of the flow reacting to the activation of the macro routes which was initially assumed to be 60%. The analyses are performed different drivers' response rates and the results are summarized in Table 9.

Table 9: Alternative routes activation for different drivers' response rates

| Drivers' response rate | Alternative routes activation |
|---|---|
| 30% | Activating the left macro route right away<br>Activating the right macro route after 1.5 hours<br>No activation of the micro route |
| 40% | Activating the left and right macro routes after 1.5 hours<br>No activation of the micro route |
| 50% | Activating the left and right macro routes after 1.5 hours<br>No activation of the micro route |
| 60% | Activating the left and right macro routes after 1.5 hours<br>Activating the micro route after 2.5 hours |
| 70% | Activating the left and right macro routes after 1.5 hours<br>Activating the micro route after 2.5 hours |

## 5. Evacuation Time for the Affected Area

According to the flood risk map, the township of San Rocco al Porto, located above the Po River, needs to be evacuated in case of a severe flood. San Rocco al Porto is a small town in the Province of Lodi in the Italian region Lombardy, located about 50 kilometers southeast of Milan and about 25 kilometers southeast of Lodi (see Figure 6).

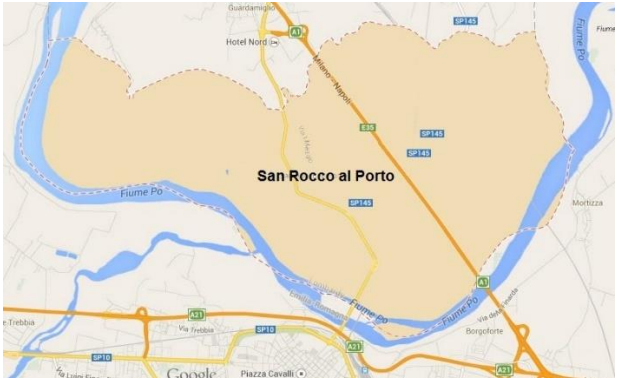

Figure 6: Location of San Rocco al Porto with respect to the Po River (source: Google Map)

According to the ISTAT 2011 census, the San Rocco Al Porto town constitutes of 48 zones with a total population of 3250 (over 1300 households) from which 66% work within the town and thus, are considered for the evacuation. A total number of 2013 vehicles are registered in the city. Assuming an even distribution of residents in each household, on average, each household consists of 1.65 evacuees and 1.55 vehicles. Accounting for the in-town population, the available number of cars will be one per household to be used for evacuation.

To model the evacuation, three main scenarios are established based on the assignment of exit points (or shelters) over the zones. In the first scenario, all the vehicles try to exit from the nearest exit point (Figure 7). This is to evaluate the situation in which there is no evacuation plan, and just the location of exit points are broadcasted to the public.

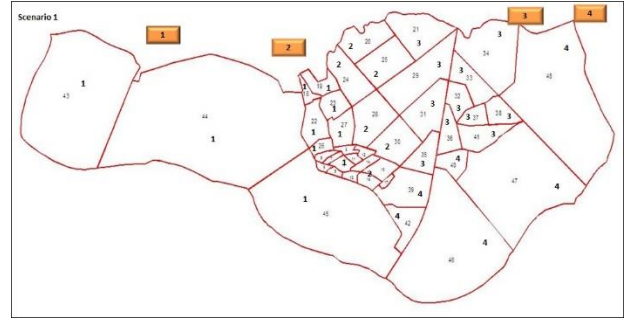

Figure 7: Evacuation scenario 1

Since the zones 5 to 17 include more vehicles to be evacuated while all are moving towards exit points 1 and 2 in the first scenario, this could be problematic. Therefore, in the second scenario, the traffic is managed to be distributed towards the 4 exit points evenly.

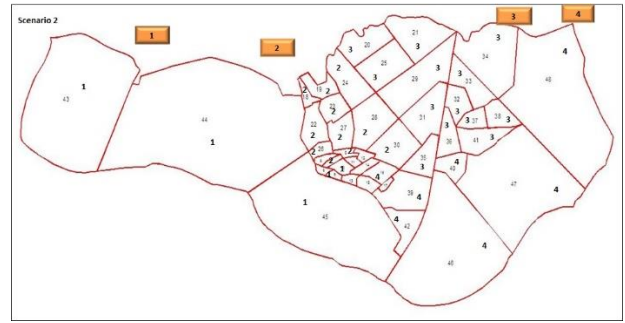

Figure 8: Evacuation scenario 2

The third scenario is developed according to the results of the second scenario, and it is considered to improve the results. Based on the results from the second scenario, evacuation toward the third and fourth exit points takes more time with respect to the other 2 exit points, therefore, in the third scenario, exit points 2 and 2 are assigned to several more zones.

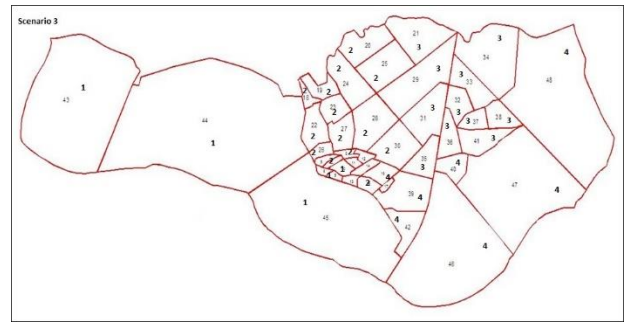

Figure 9: Evacuation scenario 3

Furthermore, for each scenario, two sets of departure time are considered. In this regard, the OD matrix is divided into four fractions; each fraction of the flow enters the network 15 minutes after the other, so the evacuation flow enters the network during 1 hour and in 4 phases every 15 minutes. The first departure time is considered as 20%, 50%, 20%, and 10%. It simulates the situation in which half of the vehicles enter the network 15 minutes after the alarm, and the rest of the flow enters early after the alarm or later. The second departure time is considered as 40%, 10%, 30% and 20%. This setting is considered to study a situation in which a specific number of residents are rather prepared for the evacuation and would leave right after the evacuation order. Table 10 lists a summary of total number of vehicles assigned to each exit point in each scenario.

Table 10: Total number of vehicles assigned to each exit point in each scenario

|  | Scenario 1 | Scenario 2 | Scenario 3 |
| --- | --- | --- | --- |
| Exit point 1 | 679 | 360 | 360 |
| Exit point 2 | 530 | 338 | 458 |
| Exit point 3 | 79 | 287 | 279 |
| Exit point 4 | 20 | 324 | 211 |
| Sum | 1308 | 1308 | 1308 |

*5.1. Results of the Evacuation*

Having the OD matrices for evacuation scenarios, the Cube Avenue Tool Pack is used to model the evacuation procedure. The results are shown in Table 11.

Table 11: Results of the evacuation scenarios

| Evacuation time (hr) | Departure Time | |
| --- | --- | --- |
|  | 1 | 2 |
| Scenario 1 | 2:52 | 2:30 |
| Scenario 2 | 1:47 | 1:46 |
| Scenario 3 | 1:35 | 1:30 |

As it is clear from the results, in all the cases the evacuation is finished in less than 3 hours; however, in the first scenario in which all the vehicles try to exit from the nearest exit point (as in lack of a detailed evacuation plan), the evacuation procedure takes about three hours which is a noticeable duration. In the second scenario in which vehicles are managed to be distributed towards the exit points, the total time is reduced favorably, and in the third scenario the results have been improved by minor interventions in the OD matrix. Comparing the results for the departure times 1 and 2, it can be observed that an optimal exit point (or shelter) allocation can compensate, to some extent, the lack of public preparedness for evacuation.

**Conclusions**

The main body of research on traffic emergency management has been devoted to challenges regarding risk assessment and emergency planning, such as traffic emergency management policy, origin-destination trip estimation, evacuation modeling, and behavior analysis. These studies have been focused on specific emergencies due to the explicit characteristics of each event. Among different modeling approaches and tools, agent-based modeling and GIS-based tools have extensive applications in transportation emergency management. These models help emergency management professionals and policy makers to evaluate different strategies during emergencies, and consequently help societies to move toward sustainability and resiliency.

In this study, a scenario involving the local flooding of the A1 motorway, one of Italy's main highways connecting north to the south, is studied. The macro itinerary, consisted of only highways, and the micro itinerary, consisted of only local roads, are established, and for different event durations and response rates, the effective route activation settings are obtained. According to the results, for drivers' response rate of 60%, when the duration of the event is less than 1.5 hours, there is no need for the activation of alternative routes; for events from 1.5 to 2.5 hours, the macro itinerary routes are required; and for events longer than 2.5 hours, the micro itinerary is also needed minimize the total travel time on the flooded route. In addition, if less than 60% of the drivers use the alternative routes, the micro itinerary would not be needed. Furthermore, the evacuation of an urban region close to the flooded area is studied considering different evacuation strategies and residents response time. The results highlight the importance of optimal exit point allocation and residents' preparedness to reduce the total evacuation time.

Acknowledgement

The study was conducted at, and the data were provided by Politecnico di Milano where the author finished his master's studies.